\documentstyle[fleqn]{article}
\input{epsf}
\def\vec#1{\mbox{\boldmath $#1$}}
\makeatletter
\def\fnum@figure{Fig. \thefigure}
\makeatother

\begin{document}

\begin{center}

{\LARGE \bf Exchange Currents \\ for Hypernuclear Magnetic Moments}

\vspace{5mm}

\large

{K. Saito}$^{\rm a}$, {M. Oka}$^{\rm a}$, and {T. Suzuki}$^{\rm b}$

\vspace{3mm}

\normalsize

${}^{\rm a}${\it Department of Physics, Tokyo Institute of Technology\\
Meguro, Tokyo 152  Japan}

\vspace{3mm}

${}^{\rm b}${\it Department of Physics, College of Humanities and Science, Nihon University \\
Sakurajosui 3-25-40, Setagaya-ku Tokyo 156 Japan}

\end{center}

\large

\begin{abstract}

The meson(K and $\pi$) exchange currents for the hypernuclear magnetic moments are calculated using the effective Lagrangian method.
The seagull diagram, the mesonic diagram and the $\Sigma^0$-excitation diagram are considered.
The $\Lambda$-N exchange magnetic moments for the ${}^5_{\Lambda}{\rm He}$, ${}^6_{\Lambda}{\rm He}$ and ${}^6_{\Lambda}{\rm Li}$ are calculated employing the harmonic oscillator shell model.
It is found that the two-body correction is about -9\% of the single particle value for ${}^5_{\Lambda}{\rm He}$.
The $\pi$ exchange current, induced only in the $\Sigma^0$-excitation diagram, is found to give dominant contribution for the isovector magnetic moments of hypernuclei with A=6.

\end{abstract}


\section{Introduction}

Physics of hypernuclei is getting to the mature phase in which not only the ground state energies and the decay rates but also the excitation spectra and their electromagnetic properties and transitions are being measured \cite{DOVER1}.
The history of nuclear physics shows us that the electromagnetic matrix elements are sensitive to the wave functions of the many body systems and therefore are important to determine its structure.
Historically, the magnetic moments of nuclei have been a driving force of
various new ideas in nuclear theory, such as the single particle picture,
the configuration mixing \cite{CONFIGMIX}, and the meson exchange currents \cite{MESONEX}\cite{CHEMTOB}\cite{MATHIOT}.
It is then expected that study of magnetic moments of hypernuclei may reveal new aspects of hypernuclear physics.

For instance, the magnetic moment may be sensitive to the spin and angular-momentum structure of hypernuclei and also to the spin dependent hyperon-nucleon interactions.
It will also probe the shell structure of the nucleon as well as the lambda in hypernuclei.
Possible mixings of $\Sigma$-hypernuclear states may also affect the magnetic moment, as is suggested in literature \cite{DOVER2}.

It is well known that mesons in nuclei make important contribution to the electromagnetic properties of nucleus.
The meson exchange currents, such as the pionic current and the delta excitation are significant in nuclear magnetic moments.
For hypernuclei, the kaon will play a similar role as the pion.
Observation of the kaon exchange current would be the first evidence of the heavy meson contributions.
Another interesting ingredient is excitation of the isovector hyperon $\Sigma$, whose mass is only 70 MeV larger than $\Lambda$.

A new aspect of hypernuclear physics is brought by the quark substructure of the hyperon.
In the naive quark picture, $\Lambda$ consists of three quarks, u, d and s.
Placing $\Lambda$ inside a nucleus may modify the structure and properties of $\Lambda$ \cite{YAMAZAKI}.
For instance, a possibility that the hyperon in nuclear matter has a different magnetic moment was suggested \cite{YAMAZAKI}.
Its confirmation will require a precise measurement of hypernuclear magnetic moments.

In the present paper, we consider the meson exchange current contributions to the magnetic moments of light hypernuclei, ${}^5_{\Lambda}{\rm He}$, ${}^6_{\Lambda}{\rm He}$ and ${}^6_{\Lambda}{\rm Li}$.
With these simple systems, we can select the various contributions for the exchange magnetic moments.
For instance, the ${}^5_{\Lambda}{\rm He}$ contains only the isoscalar $\Lambda$-N exchange current, because the ${}^4{\rm He}$ core does not have spin nor isospin.
The isovector part appears in the difference in ${}^6_{\Lambda}{\rm He}$ and ${}^6_{\Lambda}{\rm Li}$.
Their magnetic moments have contributions also from the valence nucleon in the $p$ shell.
We consider the contribution of each component separately.

The paper is organized as follows.
In Sect. \ref{se:EMMO}, we derive the exchange magnetic moment operators for $\Lambda$-N systems.
After defining the effective interaction lagrangian and the relevant coupling constants, we calculate one-kaon exchange diagrams as well as one-pion exchanges.
Excitations of the $\Sigma$($I=1$, $J=\frac{1}{2}$) are also taken into account.
We find that both the isoscalar and isovector currents arise from the kaon exchanges, while the $\Sigma$ contribution is purely isovector.

In Sect. \ref{se:MMLH}, we compute matrix elements of the exchange current operators for the light hypernuclei.
The harmonic oscillator shell model is employed for the nuclear wave functions.
The results are given in the form of simple radial integrals multiplied by the spin-flavor factor and the product of coupling constants.
Numerical values of the exchange magnetic moments are compared with the single particle magnetic moments and also with the N-N exchange current contributions in the same systems.
Discussions and conclusions are given in Sect. \ref{se:DaC}.


\section{Exchange Magnetic Moment Operators}\label{se:EMMO}


\subsection{Effective Lagrangian}

To derive the exchange currents, we employ the effective Lagrangian approach in the present study.
Here an effective meson-baryon-photon Lagrangian is defined and the relevant Feynman diagrams are computed.
The exchange current operators are derived from the static limit of the Feynman amplitudes.
Only the spatial part of the four-vector current is necessary for the magnetic moment.
The exchange currents can also be derived by using the low-energy theorem for the meson-baryon-photon vertex.
These two methods, however, are known to lead to the same results in the leading order \cite{CHEMTOB}\cite{MATHIOT}.

We adopt the interaction part of the Lagrangian density, given by
\begin{eqnarray}
&&{\cal L}_{int}={\cal L}_{{\rm strong}}+{\cal L}_{\gamma}~,
\end{eqnarray}

\noindent
where ${\cal L}_{{\rm strong}}$ is the baryon-meson interaction Lagrangian
\begin{eqnarray}
&&{\cal L}_{{\rm strong}}={\cal L}_{{\rm NN}\pi}+{\cal L}_{\Lambda {\rm NK}}
+{\cal L}_{\Sigma {\rm NK}}+{\cal L}_{\Sigma\Lambda\pi}~,
\end{eqnarray}

\noindent
and ${\cal L}_{\gamma}$ is the electoromagnetic Lagrangian
\begin{eqnarray}
&&{\cal L}_{\gamma}={\cal L}_{\Lambda {\rm NK}\gamma}+{\cal L}_{\Sigma\Lambda\gamma}+{\cal L}_{{\rm KK}\gamma}+\cdots~.
\end{eqnarray}

\begin{figure}[htbp]
        \epsfxsize=15cm
        \epsfbox{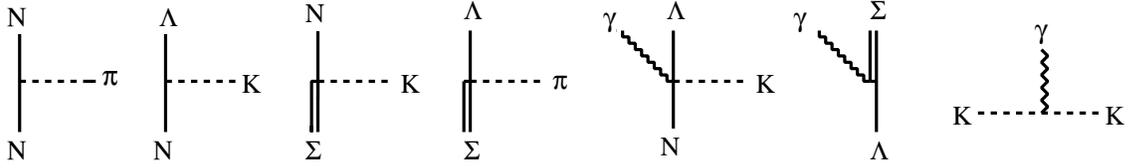}
        \caption{The interaction Lagrangian}
        \label{fig:vertex}
\end{figure}

\noindent
Each vertex is shown in Fig. \ref{fig:vertex} and is explicitly given by
\begin{eqnarray}
&&\textstyle{{\cal L}_{{\rm NN}\pi}=-g_{{\rm NN}\pi}\bar{{\rm N}}\gamma_5\gamma_{\mu}\vec{\tau}{\rm N}\cdot(\partial^{\mu}\vec{\pi})} \label{nnp}~, \\
&&\textstyle{{\cal L}_{\Lambda {\rm NK}}=-g_{\Lambda {\rm NK}}\{\bar{{\rm N}}\gamma_5\gamma_{\mu}(\partial^{\mu}{\rm K})\Lambda
+\bar{\Lambda}\gamma_5\gamma_{\mu}(\partial^{\mu}{\rm K}^{\dagger}){\rm N}\}} \label{lnk}~, \\
&&\textstyle{{\cal L}_{\Sigma {\rm NK}}=-g_{\Sigma {\rm NK}}\{\bar{{\rm N}}\gamma_5\gamma_{\mu}\vec{\tau}(\partial^{\mu}{\rm K})\cdot\vec{\Sigma}
+\bar{\vec{\Sigma}}\gamma_5\gamma_{\mu}(\partial^{\mu}{\rm K}^{\dagger})\cdot\vec{\tau}{\rm N}\}} \label{snk}~, \\
&&\textstyle{{\cal L}_{\Sigma\Lambda\pi}=-g_{\Sigma\Lambda\pi}\{\bar{\Lambda}\gamma_5\gamma_{\mu}\vec{\Sigma}
+\bar{\vec{\Sigma}}\gamma_5\gamma_{\mu}\Lambda\}\cdot(\partial^{\mu}\vec{\pi})} \label{slp}~, \\
&&\textstyle{{\cal L}_{\Lambda {\rm NK}\gamma}=-ieg_{\Lambda {\rm NK}}\{\bar{{\rm N}}\gamma_5\gamma_{\mu}\frac{1+\tau^3}{2}{\rm K}\Lambda
-\bar{\Lambda}\gamma_5\gamma_{\mu}{\rm K}^{\dagger}\frac{1+\tau^3}{2}{\rm N}\}A^{\mu}} \label{lnkg}~, \\
&&\textstyle{{\cal L}_{\Sigma\Lambda\gamma}=\frac{1}{2}g_{\Sigma\Lambda\gamma}\{\bar{\Sigma}^0\sigma_{\mu\nu}\Lambda
+\bar{\Lambda}\sigma_{\mu\nu}\Sigma^0\}F^{\mu\nu}
~~~(F^{\mu\nu}=\partial^{\mu}A^{\nu}-\partial^{\nu}A^{\mu})} \label{slg}~, \\
&&\textstyle{{\cal L}_{{\rm KK}\gamma}=ie\{(\partial_{\mu}{\rm K}^{\dagger})\frac{1+\tau^3}{2}{\rm K}
-K^{\dagger}\frac{1+\tau^3}{2}(\partial_{\mu}{\rm K})\}A^{\mu}} \label{kkg}~,
\end{eqnarray}

\noindent
where K and N are the isodoublets,
${\rm K}=$
$\displaystyle {\rm K}^+ \atopwithdelims() \displaystyle {\rm K}^0$
and
${\rm N}=$
$\displaystyle {\rm p} \atopwithdelims() \displaystyle {\rm n}$ .

We here employ the pseudovector(PV) coupling at the baryon-meson vertices and therefore the $\Lambda {\rm NK}\gamma$ vertex (seagull term) appears by the minimal substitution in the $\Lambda {\rm NK}$ vertex.
Note that other minimal coupling vertices do not contribute to the one-meson exchange currents in the $\Lambda$-N system.
The KK$\gamma$ vertex comes from the free kaon Lagrangian.
We also introduce the magnetic $\Sigma\Lambda\gamma$ effective coupling.

Although the seagull vertex disappears if we employ the pseudoscalar(PS) $\Lambda {\rm NK}$ coupling, it is well known that the pair term in the exchange current is enhanced in the PS coupling scheme and gives the same contribution as the seagull term in the static limit.
Thus to the leading order the PV and PS schemes give the same results.


\subsection{Exchange Currents}

\begin{figure}[htb]
        \epsfxsize=3cm
        \hspace{7cm}
        \epsfbox{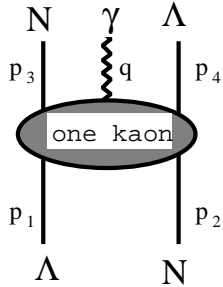}
        \caption{The momentum conventions for one-kaon exchange diagram}
        \label{fig:k-tot}
\end{figure}

As the kaon carries the strangeness, the one-kaon exchange interchanges $\Lambda$ and N i.e., $\Lambda {\rm N}\to {\rm N}\Lambda$. We here assign the kinematical variables in the exchange channel according to Fig \ref{fig:k-tot}.

\begin{figure}[htb]
        \epsfxsize=15cm
        \epsfbox{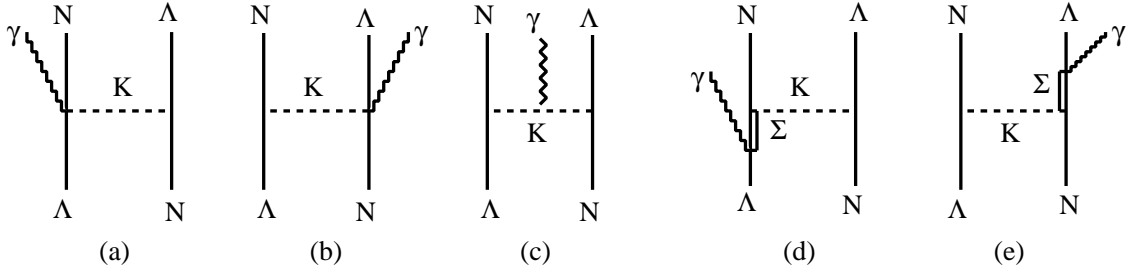}
        \caption{Diagrams for one-kaon exchange: (a),(b) seagull, (c) kaonic and (d),(e) $\Sigma^0$-excitation}
        \label{fig:k-borsig}
\end{figure}

Fig. \ref{fig:k-borsig} shows the Feynman diagrams relevant for the kaon exchange currents.
The seagull diagrams Figs. \ref{fig:k-borsig}(a) and (b) and the kaonic diagram Fig. \ref{fig:k-borsig}(c) are associated only to charged kaons.
Figs. \ref{fig:k-borsig}(d) and (e) represent $\Sigma^0$-excitation diagrams, in which both charged and neutral kaons contribute.
The corresponding exchange current operators are given after taking the static limit for the baryons.
We here neglect the mass difference of $\Lambda$ and N for simplicity, so that the exchange energy $k^0$ is set to zero.
We obtain
\begin{eqnarray}
&&\vec{\tilde{J}}^{sea}_{{\rm K}}(\vec{k}_1,\vec{k}_2;\vec{q})
=\frac{1+\tau^3_{{\rm N}}}{2}{\rm A}
\{\frac{(\vec{k}_2\cdot \vec{\sigma}_2)\vec{\sigma}_1}{\vec{k}^2_2+m^2_{{\rm K}}}
-\frac{(\vec{k}_1\cdot \vec{\sigma}_1)\vec{\sigma}_2}{\vec{k}^2_1+m^2_{{\rm K}}}\} \label{k-sea}~, \\
&&\vec{\tilde{J}}^{kao}_{{\rm K}}(\vec{k}_1,\vec{k}_2;\vec{q})
=\frac{1+\tau^3_{{\rm N}}}{2}{\rm A}
\frac{(\vec{k}_1\cdot \vec{\sigma}_1)(\vec{k}_2\cdot \vec{\sigma}_2)}{(\vec{k}^2_1+m^2_{{\rm K}})(\vec{k}^2_2+m^2_{{\rm K}})}(\vec{k}_2-\vec{k}_1) \label{k-kao}~, \\
&&\vec{\tilde{J}}^{\Sigma^0}_{{\rm K}}(\vec{k}_1,\vec{k}_2;\vec{q})
=-\tau^3_{{\rm N}}{\rm B}\{\frac{\vec{k}_2\cdot \vec{\sigma}_2}{\vec{k}^2_2+m^2_{{\rm K}}}\vec{q}\times (i\vec{k}_2+\vec{k}_2\times \vec{\sigma}_1) \nonumber \\
&&~~~~~~~~~~~~~~~~~~~~~~~~~~~~~~-\frac{\vec{k}_1\cdot \vec{\sigma}_1}{\vec{k}^2_1+m^2_K}\vec{q}\times (-i\vec{k}_1+\vec{k}_1\times \vec{\sigma}_2)\} \label{k-sig}~,
\end{eqnarray}
\noindent
where A$=e{g_{\Lambda {\rm NK}}}^2$, B$=\frac{g_{\Sigma\Lambda\gamma}g_{\Sigma {\rm NK}}g_{\Lambda {\rm NK}}}{M_{\Sigma}-M_{\Lambda}}$
, $\vec{k}_1=\vec{p}_3-\vec{p}_1$, $\vec{k}_2=\vec{p}_4-\vec{p}_2$ and $\vec{k}_1+\vec{k}_2+\vec{q}=0$.

The seagull current eq.(\ref{k-sea}) and the kaonic current eq.(\ref{k-kao}) have both the isoscalar($1$) and isovector($\tau^3_{{\rm N}}$) parts, while the $\Sigma^0$-excitation current eq.(\ref{k-sig}) gives only isovector contribution.
The latter can be understood by the fact that the $\Sigma^0$({\it I}=1) $\to$ $\Lambda$({\it I}=0)$\gamma$ vertex selects the isovector photon.

\begin{figure}[htbp]
        \epsfxsize=3cm
        \hspace{7cm}
        \epsfbox{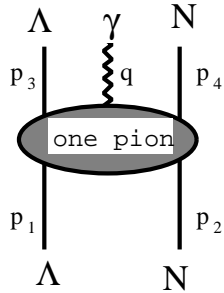}
        \caption{The momentum conventions for one-pion exchange diagram}
        \label{fig:pi-tot}
\end{figure}

\begin{figure}[htbp]
        \hspace{4cm}
        \epsfxsize=7cm
        \epsfbox{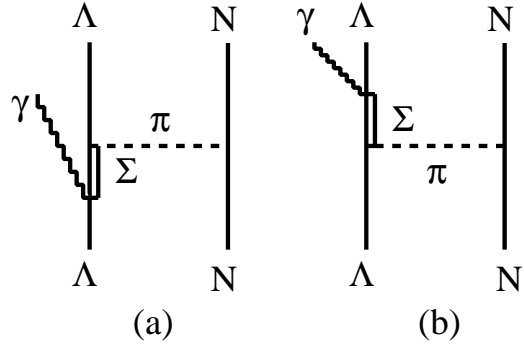}
        \caption{Diagrams for one-pion exchange: $\Sigma^0$-excitation}
        \label{fig:pi-sigma}
\end{figure}

Although the pion cannot be exchanged between $\Lambda$ and N due to the isospin conservation, the pion-exchange diagrams with $\Sigma^0$ excitation may contribute to the exchange current.
Assigning the direct-channel kinematics according to Fig. \ref{fig:pi-tot}, we calculate the Feynman diagrams, Figs. \ref{fig:pi-sigma}(a) and \ref{fig:pi-sigma}(b).
Here the diagrams such as Figs. \ref{fig:k-borsig}(a),(b) and (c) are not present because $\Lambda$ carries no isospin. The $\Sigma^0$-excitation current operator corresponding to Figs. \ref{fig:pi-sigma}(a) and (b) is
\begin{eqnarray}
&&\vec{\tilde{J}}^{\Sigma^0}_{\pi}(\vec{k}_1,\vec{k}_2;\vec{q})
=-2\tau^3_{{\rm N}}{\rm C}\frac{\vec{k}_2\cdot \vec{\sigma}_{{\rm N}}}{\vec{k}^2_2+m^2_{\pi}}\vec{q}\times (i\vec{k}_2+\vec{k}_2\times \vec{\sigma}_{\Lambda}) \label{pi-sig}~,
\end{eqnarray}

\noindent
where
C$=\frac{g_{\Sigma\Lambda\gamma}g_{\Sigma\Lambda\pi}g_{{\rm NN}\pi}}{M_{\Sigma}-M_{\Lambda}}$. Eq.(\ref{pi-sig}) has only the isovector part.


\subsection{Magnetic Moment Operators}

We transform the currents obtained in the momentum space to those in the coordinate space by the Fourier transform \cite{CHEMTOB}.
\begin{eqnarray}
&&\delta^3(\vec{x}_1-\vec{x}_3)\delta^3(\vec{x}_2-\vec{x}_4)\vec{J}^{ex}(\vec{x}_1,\vec{x}_2;\vec{q}) \nonumber \\
&&\equiv \frac{1}{(2\pi)^9}\int{\rm d}^3p_1\int{\rm d}^3p_2\int{\rm d}^3p_3\int{\rm d}^3p_4
~\delta^3(\vec{p}_1+\vec{p}_2-\vec{p}_3-\vec{p}_4-\vec{q}) \nonumber \\
&&~~~~~~~~~~~~~~~~~~~~~~~~~~~~~~~~~~~~~~\times\vec{\tilde{J}}^{ex}(\vec{k}_1,\vec{k}_2;\vec{q})
~{\rm e}^{-i(\vec{p}_1\cdot\vec{x}_1+\vec{p}_2\cdot\vec{x}_2-\vec{p}_3\cdot\vec{x}_3-\vec{p}_4\cdot\vec{x}_4)}~.
\end{eqnarray}

\noindent
Then the exchange magnetic moment operator is given by \cite{CHEMTOB}
\begin{eqnarray}
&&\vec{M}^{ex}\equiv\frac{i}{2}\vec{\nabla}_q\times\vec{J}^{ex}(\vec{x}_1,\vec{x}_2;\vec{q})|_{q=0}~.
\end{eqnarray}

We further proceed to express the $\Lambda {\rm N}\to {\rm N}\Lambda$ exchange terms into the flavor diagonal form by multiplying the exchange operator $P^f=-P^xP^{\sigma}$, where $P^x$, $P^{\sigma}$ or $P^f$ exchanges the particle numbers in the orbital, spin and flavor spaces, respectively. The spin exchange operator $P^{\sigma}\equiv \frac{1+\vec{\sigma}_1\cdot \vec{\sigma}_2}{2}$ is combined with the other spin operators.

The results are expressed as a sum of 12 operators
\begin{eqnarray}
&&\vec{M}^{ex}=\sum^{12}_{k=1}\vec{M}^{ex}_k~,
\end{eqnarray}

\noindent
where the first 8 operators, $\vec{M}^{ex}_1 \sim \vec{M}^{ex}_8$, are the contributions from the one-kaon exchange,
\begin{eqnarray}
&&\vec{M}^{ex}_1=-P^x\frac{1+\tau^3_{{\rm N}}}{2}
\frac{{\rm A}m_{{\rm K}}}{8\pi}~Y_1(m_{{\rm K}}r)~\vec{T}^{(-)}_{\Lambda {{\rm N}}}(\hat{\vec{r}})~, \\
&&\vec{M}^{ex}_2=-P^x\frac{1+\tau^3_{{\rm N}}}{2}
\frac{{\rm A}m_{{\rm K}}}{24\pi}~(1-2m_{{\rm K}}r)Y_0(m_{{\rm K}}r)~[\vec{\sigma}_{\Lambda}-\vec{\sigma}_{{\rm N}}]~, \\
&&\vec{M}^{ex}_3=-P^x\frac{1+\tau^3_{{\rm N}}}{2}
\frac{{\rm A}m_{{\rm K}}^3}{24\pi}~Y_2(m_{{\rm K}}r)~S_{\Lambda {{\rm N}}}(\hat{\vec{r}})~i[\vec{r}\times \vec{R}]~, \\
&&\vec{M}^{ex}_4=-P^x\frac{1+\tau^3_{{\rm N}}}{2}
\frac{{\rm A}m_{{\rm K}}^3}{24\pi}~Y_0(m_{{\rm K}}r)~\frac{3-\vec{\sigma}_{\Lambda}\cdot\vec{\sigma}_{{\rm N}}}{2}~i[\vec{r}\times\vec{R}]~, \\
&&\vec{M}^{ex}_5=-P^x\tau^3_{{\rm N}}\frac{{\rm B}m_{{\rm K}}^3}{4\pi}~Y_2(m_{{\rm K}}r)~\vec{T}^{(-)}_{\Lambda {{\rm N}}}(\hat{\vec{r}})~, \\
&&\vec{M}^{ex}_6=-P^x\tau^3_{{\rm N}}\frac{{\rm B}m_{{\rm K}}^3}{6\pi}~Y_0(m_{{\rm K}}r)~[\vec{\sigma}_{\Lambda}-\vec{\sigma}_{{\rm N}}]~, \\
&&\vec{M}^{ex}_7=-P^x\tau^3_{{\rm N}}\frac{{\rm B}m_{{\rm K}}^3}{4\pi}~Y_2(m_{{\rm K}}r)~\vec{T}^{(+)}_{\Lambda {{\rm N}}}(\hat{\vec{r}})~, \\
&&\vec{M}^{ex}_8=-P^x\tau^3_{{\rm N}}\frac{{\rm B}m_{{\rm K}}^3}{12\pi}~Y_0(m_{{\rm K}}r)~[\vec{\sigma}_{\Lambda}+\vec{\sigma}_{{\rm N}}]~,
\end{eqnarray}

\noindent
where A$=e{g_{\Lambda {\rm NK}}}^2$, B$=\frac{g_{\Sigma\Lambda\gamma}g_{\Sigma {\rm NK}}g_{\Lambda {\rm NK}}}{M_{\Sigma}-M_{\Lambda}}$, and
\begin{eqnarray}
&&S_{\Lambda {{\rm N}}}(\hat{\vec{r}})=3(\vec{\sigma}_{\Lambda}\cdot\hat{\vec{r}})(\vec{\sigma}_{{\rm N}}\cdot\hat{\vec{r}})-\vec{\sigma}_{\Lambda}\cdot\vec{\sigma}_{{\rm N}} \nonumber \\
&&~~~~~~~~~~=\sqrt{30}~[~{[\sigma_{\Lambda}^{(1)}\times \sigma_{{\rm N}}^{(1)}]}^{(2)}\times C^{(2)}(\hat{\vec{r}})]^{(0)}~, \\
&&\vec{T}^{(\pm )}_{\Lambda {{\rm N}}}(\hat{\vec{r}})=\{ (\vec{\sigma}_{\Lambda}\pm \vec{\sigma}_{{\rm N}})\cdot\hat{\vec{r}}\}\hat{\vec{r}}-\frac{1}{3}(\vec{\sigma}_{\Lambda}\pm\vec{\sigma}_{{\rm N}}) \nonumber \\
&&~~~~~~~~~~=-\frac{\sqrt{10}}{3}~{[C^{(2)}(\hat{\vec{r}})\times (\sigma_{\Lambda}\pm \sigma_{{\rm N}})^{(1)}]}^{(1)}~, \\
&&Y_0(x)=\frac{{\rm e}^{-x}}{x}~,~Y_1(x)=x(1+\frac{1}{x})~Y_0(x)~,~Y_2(x)=(1+\frac{3}{x}+\frac{3}{x^2})~Y_0(x)~.
\end{eqnarray}
\noindent
Here $C^{(2)}(\hat{\vec{r}})\equiv \sqrt{\frac{4\pi}{5}}~Y^{(2)}(\theta, \phi)$,
and $\vec{r}$ and $\vec{R}$ are the relative and center-mass coordinates, respectively, which are defined, assuming an equal mass for $\Lambda$ and N, by
\begin{eqnarray}
&&\vec{r}=\vec{x}_{{\rm N}}-\vec{x}_{\Lambda}~,~\vec{R}=\frac{\vec{x}_{{\rm N}}+\vec{x}_{\Lambda}}{2}~.
\end{eqnarray}

\noindent
Note that, these magnetic moment operators are nonlocal because they contain the space exchange operator $P^x$.

The remaining operators come from the $\pi$-exchange and are given by
\begin{eqnarray}
&&\vec{M}^{ex}_9=-\tau^3_{{\rm N}}\frac{{\rm C}m^3_{\pi}}{2\pi}~Y_2(m_{\pi}r)~\vec{T}_{{\rm N}}(\hat{\vec{r}})~, \\
&&\vec{M}^{ex}_{10}=-\tau^3_{{\rm N}}\frac{{\rm C}m^3_{\pi}}{6\pi}~Y_0(m_{\pi}r)~\vec{\sigma}_{{\rm N}}~, \\
&&\vec{M}^{ex}_{11}=-\tau^3_{{\rm N}}\frac{{\rm C}m^3_{\pi}}{2\pi}~Y_2(m_{\pi}r)~i[\vec{\sigma}_{\Lambda}\times\vec{T}_{{\rm N}}(\hat{\vec{r}})]~, \\
&&\vec{M}^{ex}_{12}=-\tau^3_{{\rm N}}\frac{{\rm C}m^3_{\pi}}{6\pi}~Y_0(m_{\pi}r)~i[\vec{\sigma}_{\Lambda}\times\vec{\sigma}_{{\rm N}}]~,
\end{eqnarray}

\noindent
where C$=\frac{g_{\Sigma\Lambda\gamma}g_{\Sigma\Lambda\pi}g_{{\rm NN}\pi}}{M_{\Sigma}-M_{\Lambda}}$, and
\begin{eqnarray}
&&\vec{T}_{{\rm N}}(\hat{\vec{r}})=(\vec{\sigma}_{{\rm N}}\cdot\hat{\vec{r}})\hat{\vec{r}}-\frac{1}{3}\vec{\sigma}_{{\rm N}} \nonumber \\
&&~~~~~~~~~=-\frac{\sqrt{10}}{3}~{[C^{(2)}(\hat{\vec{r}})\times {\sigma_{{\rm N}}}^{(1)}]}^{(1)}~.
\end{eqnarray}


\section{Magnetic Moments of Light Hypernuclei}\label{se:MMLH}

We now evaluate the magnetic moments of light hypernuclei, such as ${}^5_\Lambda{\rm He}$, ${}^6_\Lambda{\rm He}$ and ${}^6_\Lambda{\rm Li}$.
They are chosen as examples because their structures are relatively simple and one can see the effects of the exchange magnetic moment clearly.

We here employ the simple-minded harmonic oscillator shell model for the nucleons and also for $\Lambda$.
The magnetic moment consists of the single particle part(the Schmidt value), the exchange current between $\Lambda$ and N and the exchange and core polarization contributions of the nucleons.
We mostly concentrate on the $\Lambda$-N exchange current in the following.


\subsection{${}^5_{\Lambda}{\rm He}(J=\frac{1}{2})$}

The magnetic moment of ${}^5_{\Lambda}{\rm He}$ is given by the sum of the single particle and the two-body terms. The former is given by the magnetic moment of free $\Lambda$, assuming that all the particles are in the 0s state and therefore the total spin of the four nucleons is zero.
The two-body magnetic moment is given by the sum of the two-body terms for $\Lambda(0s_{\frac{1}{2}})$ and ${\rm N}_i(0s_{\frac{1}{2}})$ labeled by i=1,2,3,4%
\begin{eqnarray}
&&\vec{M}^{ex}({}^5_{\Lambda}{\rm He})=\sum^{4}_{i=1}\vec{M}^{ex}(\Lambda,{\rm N}_i)~.
\end{eqnarray}

\begin{figure}[htbp]
        \epsfxsize=5cm        
        \hspace{5cm}
        \epsfbox{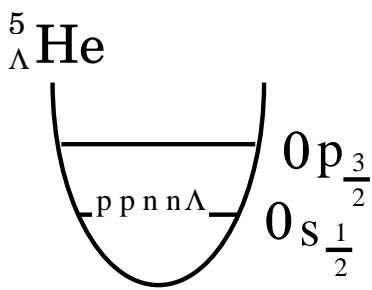}
        \caption{shell model for ${}^5_{\Lambda}{\rm He}$}
        \label{fig:shell5he}
\end{figure}

The wave function of ${}^5_{\Lambda}{\rm He}(J=\frac{1}{2})$ is given by
\begin{eqnarray}
&&\Psi ({}^5_{\Lambda}{\rm He})=\Psi (\Lambda )~\Psi ({}^4{\rm He})~,
\end{eqnarray}

\noindent
where
\begin{eqnarray}
&&\Psi (\Lambda )=\phi_{00}(\vec{x}_{\Lambda})~\chi_{\Lambda}(\textstyle{\frac{1}{2}})~, \label{psil} \\
&&\Psi ({}^4{\rm He})=\phi_{00}(\vec{x}_{{{\rm N}}_j})\sum_{s_z=\frac{1}{2} ,-\frac{1}{2}}
\sum_{t_z=\frac{1}{2} ,-\frac{1}{2}}\frac{(-)^{s_z+t_z}}{2}
~\chi_{{{\rm N}}_j}(s_z)~\eta_{{{\rm N}}_j}(t_z) \nonumber \\
&&~~~~~~~~~~~~~~~~~~~~~~~~~~~~~~~~~~~~\times \{\Psi_{others}(\vec{x}_{{\rm N}},-s_z,-t_z)\}~~~~(j=1,2,3,4)~. \label{psi4}
\end{eqnarray}

In eqs.(\ref{psil}, \ref{psi4}), $\chi_{\Lambda}$, $\chi_{{\rm N}}$ and $\eta_{{\rm N}}$ are the spin wave function of $\Lambda$, the spin and the isospin parts of the nucleon, respectively.

We take the harmonic oscillator 0s wave function for $\phi_{00}(\vec{x})$ and assume that $\Lambda$ has the same Gaussian parameter b.
\begin{eqnarray}
&&\phi_{00}(\vec{x})=R_{l=0}(x)~Y_{00}(\hat{\vec{x}})~, \\
&&~~~R_{l=0}(x)=\sqrt{\frac{4}{\sqrt{\pi}b^3}}~{\rm e}^{-\frac{x^2}{2b^2}}~. \end{eqnarray}


\subsection{A=6 isospin doublet (${}^6_{\Lambda}{\rm He}(J=1)$ , ${}^6_{\Lambda}{\rm Li}(J=1)$)}

\begin{figure}[htbp]
        \epsfxsize=5cm
        \hspace{5cm}
        \epsfbox{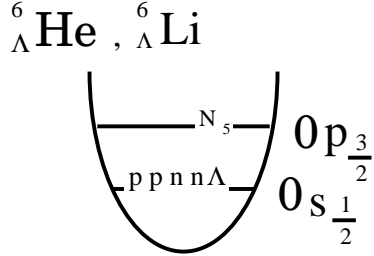}
        \caption{shell model of A=6 isodoublet}
        \label{fig:shell6}
\end{figure}

The ground state of the A=6 isodoublet has the total spin $J=1$.
We take the simplest configuration for this state, $(0s_{\frac{1}{2}})^4(0p_{\frac{3}{2}})+(0s_{\frac{1}{2}})_\Lambda$.
Then the two-body magnetic moment of the A=6 isodoublet is given by the sum of the $\Lambda$-N part and the N-N part.
We evaluate the contributions to the magnetic moment from the N-N exchange currents and core polarization for the valence nucleon according to Towner \cite{TOWNER}.
For the $\Lambda$-N part, we evaluate
\begin{eqnarray}
&&\vec{M}^{ex}({\rm A=6})=\sum^{4}_{i=1}\vec{M}^{ex}(\Lambda,{\rm N}_i)+\vec{M}^{ex}(\Lambda,{\rm N}_5)~,
\end{eqnarray}

\noindent
where ${\rm N}_5$ refers to the nucleon in the $0p_{\frac{3}{2}}$ orbit[Fig. \ref{fig:shell6}].
The shell model wave function for the $(0s_{\frac{1}{2}})^4(0p_{\frac{3}{2}})+(0s_{\frac{1}{2}})_\Lambda$ configuration is used in evaluating the matrix element of $\vec{M}^{ex}({\rm A=6})$.

\subsection{Numerical Results}

We take the values of the baryon-meson coupling constants from the Nijmegen meson exchange potential model F \cite{RIJKEN}\cite{YAMAMOTO} and the $\Sigma$-$\Lambda$ transition magnetic moment $g_{\Sigma\Lambda\gamma}$ is fixed by the $\Sigma^0 \to \Lambda\gamma$ decay rate and the quark model, which are given in Table \ref{tab:c.c.}.

\begin{table}[htbp]
\caption{Coupling constants}
\begin{center}
\begin{tabular}{ccccc}
\hline
${g_{\Lambda {{\rm NK}}}}^{-1}$ & ${g_{\Sigma {{\rm NK}}}}^{-1}$ & ${g_{\Sigma\Lambda\pi}}^{-1}$ & ${g_{{{\rm NN}}\pi}}^{-1}$ & ${g_{\Sigma\Lambda\gamma}}^{-1}$ \\
-131 MeV & 753 MeV & 201 MeV & 137 MeV & -3840 MeV \\
\hline
\end{tabular}
\end{center}
\label{tab:c.c.}
\end{table}

We use the same Gaussian parameter (b=1.35 fm) for $\Lambda$ and N, and both for A=5 and 6, for simplicity.
Other quantities needed are the $\Sigma$ mass $M_{\Sigma}$ = 1192.5 MeV, the $\Lambda$ mass $M_{\Lambda}$ = 1115.6 MeV, the nucleon mass $M_{{\rm N}}$ = 938.3 MeV, the kaon mass $m_{{\rm K}}$ = 497.7 MeV and the pion mass $m_{\pi}$ = 135.0 MeV.
Using these values, the numerical values of the magnetic moments listed in Table \ref{tab:mu} are obtained.

\begin{table}[htbp]
\caption{Numerical values of the magnetic moments}
\begin{center}
\begin{tabular}{cccc}
\hline
~ & ${}^5_\Lambda{{\rm He}}$ & \multicolumn{2}{c}{A=6 isodoublet (${}^6_\Lambda{{\rm He}}$ and ${}^6_\Lambda{{\rm Li}}$)} \\
~ & ~ & isoscalar moments & isovector moments \\
\hline
$\mu({\rm single-particle})$ & -0.613$\mu_{{\rm N}}$ & 1.090$\mu_{{\rm N}}$ & 2.377$\mu_{{\rm N}}$ \\
$\delta\mu({}^4{{\rm He}}$-$\Lambda({\rm s}_{\frac{1}{2}}))$ & 0.054$\mu_{{\rm N}}$ (-8.8\%) & -0.027$\mu_{{\rm N}}$ (-2.5\%) & --- \\
$\delta\mu(\Lambda({\rm s}_{\frac{1}{2}})$-N(p${}_{\frac{3}{2}}))$ & --- & 0.009$\mu_{{\rm N}}$ (0.8\%) & 0.044$\mu_{{\rm N}}$ (1.9\%) \\
$\delta\mu({}^4{{\rm He}}$-N(p${}_{\frac{3}{2}}))$ & --- & -0.0122$\mu_{{\rm N}}$ (-1.1\%) & -0.0536$\mu_{{\rm N}}$ (-2.3\%) \\
\hline
\end{tabular}
\end{center}
\label{tab:mu}
\end{table}

In Table \ref{tab:mu}, $\mu({\rm single-particle})$, $\delta\mu({}^4{{\rm He}}$-$\Lambda({\rm s}_{\frac{1}{2}}))$ and $\delta\mu(\Lambda({\rm s}_{\frac{1}{2}})$-N(p${}_{\frac{3}{2}}))$ denote
the single particle value, the N(s${}_{\frac{1}{2}}$)-$\Lambda({\rm s}_{\frac{1}{2}})$ exchange current contribution and the $\Lambda({\rm s}_{\frac{1}{2}})$-N(p${}_{\frac{3}{2}})$ exchange current contribution, respectively.
The last row $\delta\mu({}^4{{\rm He}}$-N(p${}_{\frac{3}{2}}))$ contains the contributions not only from the N-N exchange current but also from the core polarization.
The numbers in the parenthesis are $\delta\mu/\mu$.

As the ${}^4{\rm He}$ core has no isospin, the exchange magnetic moment for ${}^5_{\Lambda}{\rm He}$ is purely isoscalar and comes from the kaon exchanges Fig. \ref{fig:k-borsig}(a),(b) and (c).
We find that they reduce the magnetic moment by about 9\% from the single particle value, which is the magnetic moment of the free $\Lambda$.

How robust is this result?
We argue that the effect of the core polarization must be small.
The first order core polarization is forbidden as the ${}^4{\rm He}$ is $LS$ closed.
The second order core polarization is expected to be induced mostly by the tensor force.
No pion exchange, however, is allowed for $\Lambda$-N.
The $\Sigma^0$-excitation is possible, but will be suppressed by the energy denominator and also by the isospin conservation.
The two kaon exchange will be suppressed by the short range correlation.
Thus the one-kaon exchange current seems to be the only non-exotic source of the deviation from the free $\Lambda$ magnetic moment.
This is therefore a good place to look for some exotic effects, such as the quark Pauli effect for $\Lambda$, the medium modification of the $\Lambda$ magnetic moment and so on \cite{YAMAZAKI}.

For the A=6 isodoublet, apart from the sign, the isovector part is larger than the isoscalar part in both the $\Lambda$-N and N-N contributions.
Table \ref{tab:mu3} shows the contributions to $\delta\mu(\Lambda({\rm s}_{\frac{1}{2}})$-N(p${}_{\frac{3}{2}}))$ from each diagram, K-exchange(seagull and mesonic) Fig. \ref{fig:k-borsig}(a,b,c), K-exchange($\Sigma^0$-excitation) Fig. \ref{fig:k-borsig}(d,e) and $\pi$-exchange($\Sigma^0$-excitation) Fig. \ref{fig:pi-sigma}(a,b).
It is clear that the $\Sigma^0$-excitation via the pion exchange is dominant.
This result seems to be consistent with the study of the $\Sigma$ mixing contribution by Dover, Feshbach and Gal \cite{DOVER2}.
They analysed the effect of the virtual $\Sigma$ hypernuclear states to the magnetic moments of the s-shell and p-shell hypernuclei.
They found that the magnetic moment for the p-shell nuclear core is sensitive to the $\Sigma$ mixing.
Their $\Sigma$ mixing effects are, in fact, contained in the present calculation (at least partly).
Namely, the isovector $\pi$-K exchange currents with $\Sigma^0$-excitation can be regarded as the contribution of the $\Sigma$-nuclear mixing.
Then our result supports their analysis for the isovector magnetic moment, while we find that the isoscalar exchange current gives a significant effect also.
It should also be noted that the isovector part has a large (opposite) correction due to the nucleon exchange currents and the core polarization.

\begin{table}[htbp]
\caption{Contributions to $\delta\mu(\Lambda({\rm s}_{\frac{1}{2}})$-N(p${}_{\frac{3}{2}}))$ from each diagram}
\begin{center}
\begin{tabular}{cccc}
\hline
~ & \multicolumn{2}{c}{K-exchange} & $\pi$-exchange \\
~ & seagull and mesonic & $\Sigma^0$-excitation & $\Sigma^0$-excitation \\
\hline
isoscalar & 0.009$\mu_{{\rm N}}$ & --- & --- \\
isovector & 0.009$\mu_{{\rm N}}$ & -0.006$\mu_{{\rm N}}$ & 0.041$\mu_{{\rm N}}$ \\
\hline
\end{tabular}
\end{center}
\label{tab:mu3}
\end{table}

The magnetic moments of ${}^6_{\Lambda}{\rm He}$ and ${}^6_{\Lambda}{\rm Li}$ are given in Table \ref{tab:mu2}.
We see from Table \ref{tab:mu2} that the N-N contributions have the opposite sign to the $\Lambda$-N magnetic moments and therefore the net $\delta\mu$ ($\delta\mu({\rm total}))$ are 1.6\% and -1.2\% of the single particle values for ${}^6_{\Lambda}{\rm He}$ and ${}^6_{\Lambda}{\rm Li}$, respectively.
However, the single particle values of the A=6 isodoublet are larger than that of ${}^5_{\Lambda}{\rm He}$ and then the absolute values of $\delta\mu({\rm total})$ for A=6 appear in the same order as for ${}^5_{\Lambda}{\rm He}$.

\begin{table}[htbp]
\caption{The magnetic moments of ${}^6_{\Lambda}{\rm He}$ and ${}^6_{\Lambda}{\rm Li}$}
\begin{center}
\begin{tabular}{ccc}
\hline
~ & ${}^6_\Lambda{{\rm He}}$ & ${}^6_\Lambda{{\rm Li}}$ \\
\hline
$\mu({\rm single-particle})$ & -1.288$\mu_{{\rm N}}$ & 3.467$\mu_{{\rm N}}$ \\
$\delta\mu$($\Lambda$-N) & -0.062$\mu_{{\rm N}}$ (4.8\%) & 0.026$\mu_{{\rm N}}$ (0.7\%) \\
$\delta\mu$(N-N) & 0.0414$\mu_{{\rm N}}$ (-3.2\%) & -0.0658$\mu_{{\rm N}}$ (-1.9\%) \\
\hline
$\delta\mu({\rm total})$ & -0.021$\mu_{{\rm N}}$ (1.6\%)& -0.040$\mu_{{\rm N}}$ (-1.2\%) \\
\hline
\end{tabular}
\end{center}
\label{tab:mu2}
\end{table}


\section{Discussion and Conclusion}\label{se:DaC}

We have calculated the pion and kaon exchange currents between $\Lambda$ and N and evaluated their contributions to the magnetic moments of the light hypernuclei.

We find that the contribution of the isoscalar $\Lambda$-N exchange current to the ${}^5_{\Lambda}{{\rm He}}$ magnetic moment is about 9\% of the single particle (free $\Lambda$) value with the opposite sign.
Although the actual value $\delta\mu\simeq 0.05\mu_{{\rm N}}$ is small, we hope that a future "precise" measurements of the hypernuclear magnetic moment confirm our result.

The two-body contribution for the ${}^6_{\Lambda}{{\rm He}}$-${}^6_{\Lambda}{{\rm Li}}$ isodoublet is dominated by the isovector part, which is more than four times larger than the isoscalar part.
The isovector contribution mainly comes from the $\Sigma^0$-excitation through the pion exchange.
We also find that the total $\Lambda$-N two-body magnetic moments for A=6 system tend to cancel the exchange current and core polarization contributions from the nucleons.
Thus for A=6, the hypernuclear magnetic moments deviate only less than two percent from the single particle values.

Although there has been no measurement of hypernuclear magnetic moments,
several suggestions have been made for future experiments \cite{NAKAI}.
Polarized hypernuclei were already observed \cite{AJIMURA}.
A recent proposal by Nakai is to measure the magnetic moment
of ${}^5_{\Lambda}{\rm He}$ using the hyperfine magnetic field by the atomic electrons for the magnetic precession of the hypernucleus.
We hope that our prediction of the exchange current contributions to the
magnetic moments encourages such pioneering experiments.

\end{document}